# A systematic review of mechanistic models used to study avian influenza virus transmission and control

Sébastien Lambert[1*§], Billy Bauzile[1*], Amélie Mugnier[2], Benoit Durand[3], Timothée Vergne[1], Mathilde C. Paul[1]

[1] IHAP, Université de Toulouse, INRAE, ENVT, Toulouse, France

[2] NeoCare, Université de Toulouse, ENVT, Toulouse, France

[3] Epidemiology Unit, Paris-Est University, Laboratory for Animal Health, French Agency for Food, Environment and Occupational Health and Safety (ANSES), Maisons-Alfort, France

[*] These authors contributed equally to this work

[§] Corresponding author

Email addresses:

SL: sebastien.lambert@envt.fr

BB: billy.bauzile@swisstph.ch

AM: amelie.mugnier@envt.fr

BD: benoit.durand@anses.fr

TV: timothee.vergne@envt.fr

MCP: mathilde.paul@envt.fr

**ABSTRACT:** The global spread of avian influenza A viruses in domestic birds is causing dramatic economic and social losses. Various mechanistic models have been developed in an attempt to better understand avian influenza transmission and to evaluate the effectiveness of control measures. However, no comprehensive review of the mechanistic approaches used currently exists. To help fill this gap, we conducted a systematic review of mechanistic models applied to real-world epidemics to (1) describe the type of models and their epidemiological context, (2) synthetise estimated values of AIV transmission parameters and (3) review the control strategies most frequently evaluated and their outcome. Fourty-five articles qualified for inclusion, that fitted the model to data and estimated parameter values (n = 42) and/or evaluated the effectiveness of control strategies (n = 21). The majority were population-based models (n = 26), followed by individual-based models (n = 15) and a few metapopulation models (n = 4). Estimated values for the transmission rate varied substantially according to epidemiological settings, virus subtypes and epidemiological units. Other parameters such as the durations of the latent and infectious periods were more frequently assumed, limiting the insights brought by mechanistic models on these. Concerning control strategies, many models evaluated culling (n = 15), while vaccination received less attention (n = 7). According to the reviewed articles, optimal control strategies varied between virus subtypes and local conditions, and also depended on the objective. For instance, vaccination was optimal when the objective was to limit the overall number of culled flocks, while pre-emptive culling was preferred for reducing the epidemic size and duration. Earlier implementation of interventions consistently improved the efficacy of control strategies, highlighting the need for effective surveillance and epidemic preparedness. Potential improvements of mechanistic models include explicitly accounting for various transmission routes, and distinguishing poultry populations according to species and farm type. To provide insights to policy makers in a timely manner, aspects about the evaluation of control strategies that could deserve further attention include: economic evaluation, combination of strategies including vaccination, the use of optimization algorithm instead of comparing a limited set of scenarios, and real-time evaluation.

**Keywords:** avian influenza, modeling, systematic review, control strategies, disease transmission, poultry, simulations, dynamics

**TABLE OF CONTENTS**

**INTRODUCTION**

Since the emergence of the A/goose/Guangdong/1/1996 (Gs/GD) H5N1 virus in China as early as 1996 [1], highly pathogenic avian influenza (HPAI) has become a major threat to the poultry sector wordlwide [2]. Given the zoonotic potential of some variants, HPAI is also associated with serious pandemic risk and is of great concern for global public health [2]. The evolutionary and epidemiological dynamics of avian influenza viruses (AIV) took a new turn in 2014, after the emergence and spread of a new Gs/GD lineage designated clade 2.3.4.4 [3]. From 2016 to 2021, HPAI viruses clade 2.3.4.4b from the H5N8 subtype caused the largest and most severe epidemic ever reported in Europe. A new shift was observed in late spring 2021, when the HPAI H5N1 subtype became predominant. In total, 4,656 H5N8 and 4,741 H5N1 HPAI outbreaks have been reported throughout Europe from October 1, 2016 to October 1, 2022 in wild birds, poultry, and captive birds [4]. In addition to devastating losses to the poultry industry, HPAI H5Nx viruses also recently caused mass mortality events in wild birds, raising serious concern for wildlife conservation [3, 5]. The evidence of persistant HPAI virus circulation during the summer of 2021 and even more so during the summer of 2022 suggests a fundamental shift in the observed epidemiology of these viruses, with the potential for an enzootic circulation of this HPAI virus in Europe, together with repeated incursion risks via migratory birds [6, 7]. HPAI H5N1 is now also threatening North and South America, with outbreaks in wild birds and poultry.

Low pathogenic avian influenza (LPAI) viruses, which naturally circulate in wild birds and waterfowls and cause mild clinical symptoms in poultry, are also of concern since LPAI viruses of the H5 and H7 subtypes can mutate and become highly pathogenic in poultry, either shortly after the first introduction of the virus or after months or even years of undetected circulation [8].

Control measures against HPAI epidemics encompass culling of infected birds, pre-emptive culling around infected flocks, movement bans, and screening of at-risk contacts [9]. Poultry vaccination has been employed in several Asian, European and African countries [10, 11], but this strategy still faces several challenges, including difficulties in selecting vaccine strains, monitoring influenza evolution, differentiating vaccinated from infected birds as well as maintaining an appropriate vaccination coverage [11]. For this reason, poultry vaccination is currently prohibited in the European Union (EU) [12] and in the United States (US) [13]. Despite significant efforts undertaken for three decades now to limit avian influenza incursions and spread in the poultry sector, the recurrence of HPAI epidemics in a variety of epidemiological contexts (including in poultry production types considered at low avian influenza risk) raises concern about the capacity of veterinary authorities to design adequate prevention and control strategies [14].

Mechanisic models, which describe transmission dynamics by using mathematical expressions, have been widely used to evaluate the impact of existing and alternative control strategies on the spread of various animal diseases including foot-and-mouth disease [15], African swine fever [16, 17], or vector-borne diseases [18]. While numerous mechanistic models have been used to analyze avian influenza epidemics [19], an overview of evidence brought by modeling approaches for decision-making regarding surveillance and control is still lacking. Reviews conducted so far on avian influenza focused on virological and clinical aspects [20, 21], and on risk factors [22–24], including transmission routes [25, 26]. Two recent studies reviewed transmission parameter values based on experimental studies [27] and on both experimental and field studies [28], but no comprehensive analysis of mechanistic models is available to the best of our knowledge. To fill this gap, we conducted a systematic review of mechanistic models applied to avian influenza in poultry to (i) provide a description of the mechanistic models used and their epidemiological context, (ii) synthesize AIV transmission parameters, and (iii) provide insights on the impact of control measures that have been evaluated. Based on our results, we discuss future avenues and challenges for modeling AIV epidemics and evaluating control strategies.

## MATERIALS AND METHODS

The systematic review was conducted in compliance with the guidelines of Preferred Reporting Items for Systematic reviews and Meta-Analyses (PRISMA) [29].

### Search strategy

Three online databases (PubMed, Web of Science, and CAB Abstracts) were searched for relevant literature on mechanistic approaches used to study the transmission of avian influenza in domestic poultry populations. Three groups of terms were used for each database and linked with the *"AND"* Boolean; within each group, we used an association of keywords linked with the *"OR"* Boolean (Additional file 1). All searches were done in the article's title, abstract, and keywords. Articles in a language other than English were excluded. The last search was performed on October 21, 2021.

**Inclusion and exclusion criteria**

A two-step screening was performed to define the final list of articles to include in the review (Additional file 2). In the first step, title and abstracts were independently screened by two reviewers (BB and AM) based on four criteria. Articles were included if they were: (1) primary research articles on avian influenza, (2) describing a mechanistic approach for modeling the spread, (3) describing the propagation of avian influenza epidemics at the population level, and (4) focusing at least on domestic poultry population. Based on these criteria, editorials, commentaries, reviews, and perspective articles were excluded, as well as articles without any explicit description of transmission processes and articles presenting only experimental or molecular data. At this step, a conservative approach was taken where all articles selected by at least one of the reviewers were kept for the next step. The second screening step was carried out based on the full-text content. Articles were included if they met the inclusion criteria of the first screening and if the model used at least one of the following two objective: (1) estimation of model parameters using influenza epidemic data, and (2) evaluation of control strategies using a model calibrated to real-world data in the same or in a previous study. Articles solely focusing on simulated epidemics were therefore excluded. We also looked at reference lists of the included articles in order to find further articles that could have been missed in the primary search. Finally, both reviewers discussed their respective final selections until a consensus was reached on each article. In the absence of consensus, the opinion of a third reviewer (SL) was consulted.

**Data extraction and analysis**

For all selected articles, key features were systematically recorded independently by the first three authors. The information extracted included (Additional file 3): contextual information (year of the epidemic(s) studied, poultry population, avian influenza virus subtype, virus pathogenicity, geographical location, and scale), control strategies (surveillance and control measures evaluated), modeling approach (modeling aim, model paradigm, epidemiological unit, contact structure, transmission routes) and transmission parameters estimated. All extracted data were checked for consistency by the first two authors. Descriptive statistics and figures were done using R [30].

**RESULTS**

**Included articles and epidemiological characteristics**

Based on an initial search on April 24, 2019 and a search update on October 21, 2021, the search query yielded 2,669 articles, of which 1,041 were duplicates (Figure 1). Of the 1,628 remaining articles, 1,353 and 234 articles were removed after the first and second screening, respectively. Four additional articles that met the inclusion criteria were identified by checking the references of the articles that passed both screenings. As a result, a total of 45 articles were included in the review [31–75] (Figure 1; Additional file 4).

The epidemiological context of the models presented in the included articles varied in terms of pathotype (LPAI or HPAI), subtype, and location (Table 1, Figure 2). The vast majority of articles focused on HPAI (42/45, 93%), with few studies on LPAI (6/45, 13%, Table 1).

The majority of the articles (20/45, 44%; Table 1, Figure 2) focused on the HPAI H5N1 subtype. Although HPAI H5N1 has spread globally since its emergence in 1996, it is noteworthy that models were applied to only a limited number of countries, mainly in Asia (n = 14) with Thailand, Vietnam, Bangladesh, India, Indonesia and South Korea. Four articles analyzed HPAI H5N1 spread in Africa (Nigeria and Egypt), one in Europe (Romania), and two at the global scale.

The second most studied subtype was HPAI H7N7 (7/45, 16%; Table 1, Figure 2). All seven articles, including the earliest modeling article published in 2004 [65], investigated the 2003 HPAI H7N7 epidemics in The Netherlands.

Four articles (9%) focused on LPAI and HPAI H7N9 in China [72–75]. Since its emergence in early 2013, the influenza A H7N9 virus has caused six epidemic waves, with over 1,500 human infections in mainland China [73, 76]. In addition to zoonotic concerns, the H7N9 virus also raised major threats to the poultry industry when it mutated into a highly pathogenic form in 2017 in Guangdong province before spreading rapidly to the whole country [77].

Other past HPAI epidemics analyzed so far included H5N8 in South Korea, France and The Netherlands (n = 5, 11%), H5N2 in the United States (n = 4, 9%), H5N6 in South Korea and the Philippines (n = 4, 9%), H7N3 in Canada (n = 1, 2%), and H7N1 in Italy (n = 1, 2%). Appart from LPAI H7N9 in China (n = 4, 9%), only two other studies focused on LPAI viruses (Table 1): LPAI H5N2 in the United States (n = 1, 2%) and LPAI H7N3 in The Netherlands (n = 1, 2%).

**Modeling approaches**

In accordance with our inclusion criteria, the most frequent modelling objective was to estimate model parameters (42/45) (see Parameters estimations), followed by the evaluation of control strategies (21/45) (see Mitigation strategies), with 18 articles doing both. Three different modeling approaches (Box 1) were used in the selected articles: population-based models (PBM, 26/45 articles), individual-based models (IBM, 15/45 articles), and metapopulation models (MPM, 4/45 articles; Table 2). All PBMs assumed homogenous mixing between epidemiological units (i.e. where one epidemiological unit could contact any other unit in a population randomly with equal probability). IBMs assumed that effective contacts depended on the distance between two epidemiological units: either contacts occurred with equal probability but only with units within a given radius, or contacts could happen with any other unit but with a probability that decreased with increasing distance. MPMs models assumed homogenous mixing between epidemiological units (birds or farms) within each subpopulation (see Box 1), as well as distance-dependent contacts between different subpopulations (farms or geographical areas).

Three types of epidemiological units were considered: individual birds, poultry farms, or geographical areas defined by administrative boundaries (e.g. villages, districts, counties, countries…). When considering individuals as epidemiological units (17/45 articles, Table 2), health statuses of birds were classicaly defined as susceptible (S), exposed – infected but not yet infectious (E), infectious (I), recovered (R) or dead (D). Most frequently used were SID (n = 9) and SEIRD (n = 5) models. When considering farms as epidemiological units, the whole farm was considered as exposed and then infectious after the onset of infection (i.e. neglecting within-farm dynamics). Similarly, for administrative units, the whole area was considered as exposed when at least one outbreak (e.g. one infected farm) occurred, and then the whole area became infectious at the end of the latent period. At these levels (farms, areas), the recovered state was not considered: after being infectious, the whole epidemiological unit was either considered as completely depopulated, or completely susceptible again. At the farm and administrative unit levels (28/45, Table 2), the most frequent models were SEID (n = 15) and SID (n = 9).

Most PBMs that are present in the review (9/26; Table 2) were applied at a subnational scale and considered zoonotic transmission between individual birds and humans [72–75], transmission between poultry farms [64, 65, 68], or transmission between poultry populations pooled at the administrative unit level [33, 45]. Developed at a national scale, eight articles used PBMs that considered as epidemiological units either farms [35, 37, 39, 44], administrative units [47, 50], or individual birds [49, 56]. Seven articles used PBMs to investigate within-farm transmission dynamics [31, 54, 55, 59, 61, 63, 69]. Finally, two articles used PBMs to model HPAI virus transmission between individual birds and humans on a global scale [40, 41].

Articles with IBMs were mostly applied at the subnational scale (11/15; Table 2) to investigate the transmission process between farms [42, 43, 51–53, 57, 60, 62, 66] or between administrative units [32, 36, 52]. The remaining four articles that used IBMs were applied at the national scale to investigate the transmission process between farms [38, 46, 67, 71] or between administrative units [46]. Some authors evaluated in the same study separate epidemiological units [46, 52].

Finally, the four articles with MPMs investigated the transmission process between individuals [48, 70] or between farms [34, 58] (Table 2).

Data on poultry species explicitly included in the model was specified in only 20 out of the 45 articles. It included mainly chickens (n = 17), ducks (n = 9), and turkeys (n = 6). Three articles included other species, such as geese, quails, and ostriches [32, 60, 62]. Overall, poultry populations were usually considered as one homogeneous population, even though the information may be present in the demographic data for each poultry species. There were a few exceptions that distinguished backyard from commercial farms [32, 64, 68] or that incorporated the heterogeneity in transmission between the host species studied and sometimes estimated their relative contribution to transmission [32, 34, 38, 58, 60, 62].

Similarly, heterogeneity in transmission routes were rarely considered. Most transmission models considered that AIV transmission occured only via poultry-to-poultry transmission, with only 11 of the 45 considering transmission via other routes (e.g. contacts with wild birds or contaminated environment). For example, some authors modeled explicitly the environment [41, 73, 74] or the disease dynamics in wild birds [32, 41, 45, 48], while others considered a constant parameter to implicitly capture other undefined transmission sources (e.g. infectious backyard farms or wild birds, long-distance movements of infected birds or contaminated equipment) [34, 42, 43, 53, 60].

Finally, zoonotic transmission was considered in 10 articles [40–42, 48, 49, 56, 72–75].

**Parameters estimations**

The between-individual transmission rate $\beta$ was systematically estimated in all eight articles considering within-farm transmission (seven PBMs and one MPM; Table 3). Although all eight models assumed homogenous mixing and frequency-dependent contact rates, the values of the between-individual transmission rate ranged from 0.5 to 34.4 per day (Table 3). Accordingly, values for the between-individual basic reproduction number $R_0$ ranged widely, from 2.18 to 17.5 (Table 3). The most common parameters other than $\beta$ and $R_0$, i.e. the average durations of the latent and infectious periods and the case fatality risk, were most often fixed using published values from infection experiments (Additional file 5). The case fatality risk and the duration of the average latent period were estimated in only one study [61], and the average duration of the infectious period was estimated in two studies [61, 63]. The average duration of the latent period was always lower than two days, irrespective of the subtype and the pathogenic form, and the average duration of the infectious period ranged from 1 to 15 days (Additional file 5). The case fatality risk ranged from 20% to 100% for HPAI viruses, while it was very low (0 – 1%) for LPAI viruses. Four of the eight articles considering within-farm transmission also estimated the time of virus introduction in addition to the above-mentionned parameters [55, 59, 61, 63]. All parameter values in within-farm transmission models of HPAI viruses were estimated using daily recorded mortality data, while egg production data [63] or diagnostic tests results data [55] were used for LPAI viruses (Table 3).

Similarly, the between-farm transmission rate $\beta$ and the between-farm $R_0$ were the most commonly estimated parameter in the 23 articles studying between-farm transmission (Table 3). Although the transmission rates were less comparable because of differin assumptions (e.g. frequency- vs density-dependence – Box 1), the corresponding estimates of the $R_0$ ranged widely, from 0.03 to 15.7 (Table 3). The average duration of the infectious period at the farm level was the third most frequently estimated parameter, and ranged quite widely depending on epidemiological settings and virus subtypes. However, the smallest estimated value was 6.4 days, indicating rather long infectious periods at the farm-level across studies (Additional file 6). The average durations of the other health states (i.e. latent period – from onset of infection to onset of infectiousness, incubation period – from onset of infection to notification, and clinical period – from notification to depopulation) were more often assumed than estimated, based on experimental data or field observations (Additional file 7).

Additionally, 14 (13 IBMS and one MPM) of the 23 articles modeling between-farm transmission used spatial transmission kernels to describe how the relative risk of transmission between farms changed with the distance between a susceptible farm and an infectious farm. The resulting spatial transmission kernels are illustrated in Figure 3, except for Seymour et al. [71] who used a non-parametric transmission kernel that could not be reproduced. Two studies used a step function where the relative risk was one below a certain distance threshold, and zero otherwise [38, 60]. Le Menach et al. [66] estimated three transmission rates for pre-determined ranges of between-farms distances (less than 3 km, between 3 and 10 km, and more than 10 km), which we converted to a step-function transmission kernel (Figure 3). The other studies used parametric kernels, with three different functions (Additional file 7): a Pareto distribution [42, 43], a power-law function [46, 51, 52], or a logistic expression [53, 57, 62, 67, 70]. The parameters of these kernels were all estimated (Additional file 7), except for Backer et al. [70] and Hill et al. [43], which used the same parameter values as Boender et al. [67] and Hill et al. [42], respectively. Rorres et al. [51, 52] and Pelletier et al. [46] estimated large values of $\rho$ and very small values of $\delta$ (Additional file 7), thus producing a step-like function where the relative risk of transmission between farms decreased sharply to very small values (Figure 3). However, in these articles, the force of infection accounted for the number of birds in the susceptible and infectious farms, meaning that the probability of transmission between farms remained substantial even at high distances. Hill et al. [42] also found a rapidly decreasing probability, but with long right-tails depending on the epidemic used to parameterise the kernel (Figure 3 and Additional file 7). Similar values of $\alpha$ were found in Boender et al. [67], Dorigatti et al. [62] and Bonney et al. [53], thus producing similar shapes, but different values of the half-kernel distance $r_0$, with local transmission being more important in Boender et al. [67] and Dorigatti et al. [62], while transmission at moderate distances remained significant in Bonney et al. [53]. Conversely, Salvador et al. [57] found a similar half-kernel distance than Dorigatti et al. [62], but a smaller shape parameter $\alpha$, indicating a larger role of long-distance transmission (Figure 3 and Additional file 7).

Finally, the remaining 15 articles considered transmission within- and/or between-administrative units, with individuals or administrative units as epidemiological units (i.e. not modeling farms).

Eleven (73%) estimated the transmission rate and the $R_0$, three (20%) estimated the average duration of the infectious period, and two (13%) estimated the average duration of the incubation period and spatial kernel parameters. However, because of the variations between epidemiological units and spatial scales considered, limiting comparisons and the possibility to use values for different settings, we did not report the estimated parameter values here (but see Additional file 3).

**Mitigation strategies**

Out of the 45 selected articles, 21 evaluated the effectiveness of control measures to contain AIV outbreaks through numerical simulations (Table 4). The three modeling approaches were represented with 12 studies using IBMs, 5 using PBMs and 4 using MBMs. Most articles (15/21) have studied the effect of control measures based solely on epidemiological parameters in the poultry population (size and duration of the epidemic, $R_0$, number of culled flocks…). The economical impact of implemented measures was considered in only three articles [34, 70, 71]. Finally, in three studies [49, 73, 75], the authors estimated the effectiveness of mitigation strategies in poultry for controlling the disease in humans (e.g. reduction of the number of infected humans).

The most commonly evaluated strategy was poultry culling (15 articles, Table 4), either in the form of reactive culling (RC: culling of infected flocks, 5 articles) or pre-emptive culling (PEC: culling of at-risk flocks not necessarily infected, 12 articles), with two articles evaluating both. RC was most of the time included in the model's baseline mitigation stategy without evaluating its effectiveness on its own [32, 34, 36, 43, 45, 46, 48, 53, 57, 58, 60, 62, 66, 67, 70, 71, 75]. As an exception, Lee et al. [49] showed that reducing the number of infectious poultry led to a substantial reduction in the number of H5N1 infections in humans compared to a scenario without reactive culling. However, the impact of the timeliness of RC was evaluated in four articles, which all consistently showed a strong positive impact of this parameter on the effectiveness of RC [45, 48, 60, 66]. Andronico et al. [60] for example, showed that a reduction of the delay between detection and culling (from 5 to 2 days) reduced the expected size of the outbreak in poultry by a factor of 2.

PEC was implemented as the culling of flocks in a given radius around detected infected flocks, which was below 10 km in most articles [34, 43, 46, 57, 58, 60, 62, 66, 67, 70, 71], except for Kim et al. [38] which evaluated radii up to 56.25 km. All articles reported the effectiveness of PEC on reducing the number of infected flocks and the duration of the epidemic, with increasing effectiveness as the culling radius increased. However, the economic costs of the epidemic were also the highest for the highest PEC radii [70, 71].

Moreover, the improvement brought by increasing the preventive culling radius saturated rapidly when considering limited culling capacity [43, 46, 58, 60, 66, 70, 71]. This is because further increasing the culling radius did not change the number of culled flocks once the maximum culling capacity was reached. Limited culling capacity was introduced to represent limited ressources and logistical difficulties presented by mass cullings over a short period of time, and were either modeled as a number or proportion of preventively culled farms around each infected farm [60, 66], as a maximum number of farms culled at each time step [43, 46, 70, 71], or as a nonlinear culling rate depending on the number of reported farms [58]. Increasing the culling capacity substantially increased the effectiveness of PEC on the epidemic size and duration [43, 70].

The effectiveness of PEC came at the cost of culling more flocks. In most cases, the total number of culled flocks (by RC and PEC) increased as the PEC radius increased, meaning that RC without PEC would be preferred if the objective was to reduce the number of culled flocks instead of reducing the epidemic size [34, 43, 60, 62, 70, 71]. In some cases, however, the total number of culled flocks was non-monotonic as the PEC radius increased [38, 57, 58]. The total number of culled flocks first decreased, which could be explained by the number of infected flocks decreasing faster than the number of pre-emptively culled flocks increased. Then, the total number of culled flocks increased again as there were more and more pre-emptively culled flocks. Therefore, there was a minimum number of culled flocks at an optimal PEC radius, which ranged from 1 to 18.75 km depending on the epidemiological setting [38, 57, 58]. In Lee et al. [58], this non-monotonic behavior was only observed in areas with $R_0$ above one, but not in areas with $R_0$ below one.

Vaccination strategies were less frequently evaluated than culling strategies (seven articles; Table 4). Vaccination was applied either in a radius of 1 to 10 km around infected flocks [43, 46, 70] or in a variable proportion of the total poultry population [34, 36, 46, 56, 73]. One article evaluated the impact of a vaccination strategy that was applied in real life and not as a theoretical scenario [36]. The results showed that the country-wide vaccination strategy implemented in Vietnam from 2005 with a ~60% coverage allowed to substantially reduce the transmissibility of HPAI infection. However, this was coupled with a substantial increase of the time from infection to detection, possibly because of lower levels of mortality and symptomatic infections making outbreaks harder to detect [36]. When considering hypothetical vaccination strategies, the results showed the effectiveness of vaccination from an epidemiological point of view in most articles, although ring vaccination was found ineffective in some cases [43, 46]. Pelletier et al. [46] showed that ring vaccination was less effective than countrywide vaccination with 92-97% coverage. The effectiveness of coutrywide vaccination was further improved when the order in which premises were vaccinated was determined by known risk factors, such as flock size or proximity to an infected flock [46]. As for PEC, increasing the vaccination radius size brought limited improvements when considering limited ring vaccination capacity [43]. Furthermore, increasing the capacity was not as effective for ring vaccination as it was for PEC [43, 70].

Vaccination was compared to PEC in four articles [34, 43, 46, 70]. Their respective effectiveness depended on the objective: vaccination was more effective in reducing the number of culled flocks, while PEC was more effective in reducing epidemic size and duration [34, 43, 46, 70]. In Pelletier et al. [46], country-wide vaccination was the most effective on reducing the number of culled flocks, while at the same time having similar effectiveness than PEC in reducing the epidemic size. However, the control effort required by these two strategies was substantially different, with 92 to 97% of flocks that were vaccinated compared to only 17-27% that were culled [46].

Only two studies compared the relative costs of PEC and vaccination strategies [34, 70]. Backer et al. [70] found that although the duration and size of the epidemics were lower for PEC than for ring vaccination for a same radius (3 km), the total costs (direct costs such as compensation of culled poultry, costs of culling, cleaning and distinfecting, costs of vaccine doses and vaccination, and surveillance costs, as well as indirect costs such as lower prices for eggs and slaughtered poultry, and loss of revenue because of business interruption) of both strategies differed only marginally. Retkute et al. [34] showed that the choice of the strategy depended on the relative costs of preventive culling and vaccinating compared to the economic impact of having infection in a flock (e.g. costs of reactive culling, compensation paid for destroyed animals, or costs associated with potential zoonotic transmission to poultry professionals). If the cost of culling was low compared to the cost of an infected flock, PEC was preferred regardless of the cost of vaccination. If the costs of culling and vaccinating were both high compared to the cost of an infected flock, then neither strategy were preferred and RC only minimised the total costs of the epidemic. However, if the cost of culling was high compared to the cost of an infected flock but the cost of vaccination was low, then vaccination was preferred [34].

Eight articles evaluated the effectiveness of surveillance strategies (Table 4). Andronico et al. [60] demonstrated that the surveillance zone implemented as part of the mandatory EU strategy during the 2016-2017 H5N8 epidemic in France was effective in reducing transmission, while Walker et al. [32] demonstrated the effectiveness of intensive surveillance campaigns in reducing the delay between infection and report during the 2004-2005 H5N1 epidemic in Thailand. The other six articles all showed that improving surveillance to reduce the delay between infection and detection had a large effect on the epidemic [36, 38, 43, 48, 57, 66]. For instance, reducing the average delay between infection and detection by 35% reduced the epidemic size by 66.8% in Walker et al. [36]. Hill et al. [43] even concluded that improving surveillance was a more effective strategy than PEC or ring vaccination in most cases.

Various other strategies (e.g., movement bans, biosecurity measures, closure of live poultry markets) were evaluated in seven articles (Table 4). Ban of restocking on emptied farms during an epidemic, either because they completed a production cycled or because they were culled, was often assumed but its effectiveness was evaluated only in Dorigatti et al. [62]. They showed that, if the ban on restocking had not been implemented, the epidemic size would have been 155% times higher on average, demonstrating the importance of this mitigation measure [62]. During the H5N2 epidemic in Minnesota (US) in 2015, a strategy called "early marketing" was used, which consisted in sending flocks to abattoirs earlier than the normal schedule date to reduce the density of poultry farms [53]. This strategy allowed to sufficiently reduce the density to decrease the reproduction number below one in a high-density area, therefore reducing the spread of the virus [53]. Therefore, it could represent an interesting alternative or complement to the classical PEC strategy.

Finally, across mitigation strategies, implementing measures earlier during an epidemic was always better [62, 66, 75]. When considering the impact of mitigation strategies on zoonotic transmission to humans, it was demonstrated that combination of measures in poultry with measures in humans were better than measures in poultry alone [49, 73]. Interestingly, as implementing a single strategy at a high effective level may be difficult, combining two strategies at lower levels could be easier to implement while being as or even more effective, thus providing an interesting alternative [49].

**DISCUSSION**

Epidemiological models with a mechanistic approach have been increasingly recognized as valuable for analyzing and developing control strategies for infectious disease outbreaks [78], including avian influenza [19]. When they are built with a continuous dialogue between decision-makers and the infectious disease community, and by drawing on real-world data, epidemiological models can provide insights into new approaches to prevention and control, which can help shape national and international public health policy.

The results presented in this study summarise the state of the art in mechanistic modeling applied to real-world epidemics of low- and highly-pathogenic avian influenza in poultry. Our objective was to synthesize the current knowledge on AIV transmission parameters, provide insights on the impact of control strategies, and discuss future avenues for modeling AIV transmission. We limited our analysis to articles that showed a validated model with epidemic data to ascertain the articles' relevance for this review. The search was restricted to three online databases for articles in English based on a selected set of keywords. Therefore, it is possible that a few relevant publications in the grey literature, without the selected keywords or in another language than English were missed. However, we are confident that this analysis provides an accurate literature synthesis.

**Epidemiological insights**

*Transmission mechanisms*

As illustrated in Figure 3, different transmission models used distance-based kernels to study AIV spread between poultry farms. The estimated kernel parameters (Additional file 7) suggest that most AIV transmission happened at a short to moderate distance range, irrespective of the subtype or geographical location. This is in line with what has been observed in observational studies [23, 79]. However, it does not mean that long-distance transmission is impossible. For example, the transmission models used in Bangladesh [42] and the Philippines [57] were long-tailed, indicating that some transmission events were often sourced over long distances. Differences in spatial transmission kernels were observed between epidemics in different countries, but changes could also happen between different epidemics of the same subtype in a given country as highlighted in Hill et al. [42] or even between different waves in a given epidemic as suggested in Bonney et al. [53]. Such differences could arise from contrasted poultry systems, distinct characteristics of different AIV subtypes, or specific AIV control strategies.

Poultry farming systems vary widely regarding species raised, farm sizes, and management systems [80]. This heterogeneity is likely to play a role in the epidemic dynamics, but was rarely accounted for explicitly. However, when species heterogeneity was explicitly modeled, it was possible to determine the species respective contributions to transmission, which appeared to depend on the virus subtype. For example, Andronico et al. [60] explicitly modeled galliformes (e.g. chickens) and palmipeds (e.g. ducks and geese) to account for their contribution to the 2016 – 2017 H5N8 propagation dynamics in France. They were able to show that palmiped farms were 2.6 times (95%CI: 1.2-10) more susceptible and 5.0 times (95%CI: 3.7-6.7) more infectious than galliform farms. Similar results were found in the Republic of Korea for the 2016 – 2017 H5N8 spread, with transmissibility between duck farms being as much as 1.6 times higher than either between chicken farms or between different farm types [58]. In contrast, chicken farms were up to 10.7 times more infectious than palmiped farms during the 2004 HPAI H5N1 epidemics in Thailand [32].

Similarly, heterogeneity between management systems was explicitly accounted for in three articles by considering separate compartments for commercial and backyard poultry farms [32, 64, 68]. Both Bavinck et al. [68] and Smith and Dunipace [64] found that backyard farms played a limited role in AIV transmission, with reproduction numbers below one for transmission between backyard farms and between commercial and backyard farms (note that Walker et al. [32] used a common compartment for backyard poultry and wild birds, preventing comparison). However, accounting for backyard farms remains necessary to accurately estimate the effort required for control strategies targeting commercial farms only [64].

Wild birds were sometimes included to account for sources of infection other than domestic poultry farms. Different approaches were used, either by implementing a constant parameter fitted to the epidemic data and implicitly considering various sources of transmission (including wild birds) [34, 42, 53, 60], by including seasonal introductions into poultry farms via migrating birds [41] or by considering a specific compartment for the number of wild birds in the vicinity of domestic farms [32, 45, 48]. When the relative role of wild birds and domestic poultry was assessed, commercial poultry farms were found to be the main sources of infection [41, 45, 60]. For instance, Andronico et al. [60] quantified that wild birds and backyard poultry only accounted for 11% of transmission events. However, wild birds may be playing an important role in introducing and re-introducing the virus to the local environment [41, 45].

*Parameter estimates*

Two recent studies reviewed values of the transmision rate, the basic reproduction number, and the average durations of the latent and infectious periods estimated from experimental studies [27, 28]. In addition, Kirkeby and Ward [28] included estimates from field data, but only if no disease control strategies had been implemented when the data used to perform the estimation was collected. Our review of parameter values therefore differed from these previous studies, as we excluded studies based on experimental infections and we included all estimates from field data, including those from periods with active disease control strategies. This was in line with our objective, which was to review both parameter values and effectiveness of disease control strategies estimated using mechanistic models of real-life AIV epidemics.

The estimates presented in the selected studies for the current review were from a wide range of geographical regions and for various subtypes. This will undoubtedly improve risk analysis and the creation of future models by constructing more realistic simulation studies. While some insights may be gained into the virus spread dynamics in these areas, caution should be taken to extrapolate parameters from one region to the next as disease dynamics depend on numerous factors, including local conditions, topology, farm density, industry system, and many other factors [81]. As an example, one of the estimated parameters is the $R_0$, the average number of secondary cases produced by an average infectious epidemiological unit (individuals, farm, or administrative unit), should all other epidemiological units be susceptible. Estimated values of this parameter varied substantially depending on countries and subtypes, e.g between 2.18 and 17.5 for within-farm transmission, or between 0.03 and 15.7 for between-farm transmission.

*Control strategies*

About half of the included articles evaluated control strategies (21/45). Overall, both reactive (RC) and pre-emptive (PEC) culling remained the most studied control measures, reflecting the predominant availability and usage of these control measures in real-life situations. In contrast, vaccination was more rarely evaluated. The use of vaccination as a routine component of the control strategy is currently prohibited in the EU [12] and the US [13] because of the trade implications, but is now debated in Europe because of the recent recurring and devastating epidemics [3, 13]. Indeed, vaccination has been shown to be an efficient tool to curb the impact of AIV spread elsewhere: for instance, vaccination in the poultry sector in China against the zoonotic HPAI H7N9 has decreased the number of cases in both poultry and humans [3, 13, 73]. The effectiveness of control measures is likely to depend on the virus subtype and on local conditions, such as farm density, distribution of farm types (backyard/commercial) and of poultry species. For instance, the optimal radius minimizing the number of preventively culled flocks varied between locations and epidemic years [38, 57, 58]. Similarly, the optimal strategies minimizing the costs of AIV epidemics depend on the relative costs of various control measures [34], which may vary over time and between countries. Therefore, it is critical that the evaluation of control strategies using modeling approaches are tailored to local contexts that are contemporary to the situation at stake, whenever possible.

Nonetheless, we identified two general findings on the evaluation of control strategies. First, the optimal control measure depended on the objective: often, the best strategy was different whether the objective was to reduce the epidemic duration and size or whether the objective was to reduce the number of culled flocks [34, 43, 46, 70]. In particular, vaccination was often the strategy associated with the lowest number of culled flocks, thus possibly reducing the costs of an epidemic (e.g. median of 278 culled farms for the baseline scenario *vs* only 163 culled farms in the same scenario but with ring vaccination [70]). However, vaccination strategies may incur other costs by increasing epidemic durations (e.g. median of 67 days for ring vaccination compared to a minimum of 26 days for ring culling [70]) or by disrupting international trade. Note that losses due to restriction on exportations were not included in the three models that compared control strategies based on economic costs [34, 70, 71]. Therefore, a careful definition of the objective and of the costs associated with each measures are necessary before evaluating control strategies.

Second, early implementation of control measures supported by effective surveillance was always beneficial. For instance, reactive culling of infected flocks soon after their detection had a substantial impact on the course of AIV epidemics [45, 48, 60, 66]. Similarly, improving surveillance to reduce the delay between infection and detection/report, thus allowing earlier implementation of various strategies (e.g. reactive culling of the infected flock, pre-emptive culling around it…), substantially reduced the impact of AIV epidemics [32, 36, 38, 43, 48, 57, 66]. Moreover, the effectiveness of control strategies increased as the delay between the first detected outbreak and the first implementation of control measures decreased [62, 66, 75]. For the 1999-2000 HPAI H7N1 epidemic in Italy, the number of outbreaks would have been reduced from 385 to 222 on average if preventive culling had been implemented 20 days after the epidemic started rather than 54 days [62]. All these results highlight the importance of epidemic preparedness, with effective surveillance systems and quick responses of policy makers and veterinary services.

*Different insights for different modeling approaches?*

AIV transmission has sometimes been evaluated for the same epidemics in the same countries but with different modeling approaches (Table 1). That was the case for instance for the 2003 HPAI H7N7 epidemics in The Netherlands [65–68, 70, 71] and for the 2004 HPAI H5N1 epidemic in Thailand [32–34]. In these epidemics, different modeling approaches were applied to the same datasets with varying levels of details, providing varying parameter estimates. The 2003 HPAI H7N7 epidemics in The Netherlands was by far the most studied (Table 1) and provided a good case study of how various modeling approaches influence parameter estimations and recommandations to policy makers. Two population-based models neglected the spatial aspect of between-farm transmission [65, 68]. Although Bavinck et al. [68] accounted for backyard flocks and neglected temporal changes in transmission rates and infectious periods during the study period (22 February-3 April 2003), they estimated a between-farm reproduction number ($R_0$) of 1.33 (Table 4), quite similar to the 1.2 (95% CI: 0.6-1.9) estimates for the same area between 1 March 2003 and 3 April 2003 in Stegeman et al. [65]. Bavinck et al. [68] had the additional advantage of estimating the contribution of backyard and commercial flocks to transission, and showed that culling backyard flocks may not be necessary as they only played a minor role [68]. Interestingly, estimates of the between-farm $R_0$ were similar between a non-spatial PBM [65] and a spatial IBM [66]. In Stegeman et al. [65], the estimates were 6.5 (95% CI: 3.1-9.9) during the first phase of the epidemic and 1.2 (95% CI: 0.6-1.9) during the second phase of the epidemic (Table 4), whereas they were 5.2 (95% CI: 4-6.9) for the first phase and 1.5 (95% CI: 1-2.5) for the second phase in Le Menach et al. [66]. Both studies concluded that the implemented control measures (i.e. RC, movement bans, contact-tracing, PEC within a 1-km radius) were not sufficient to halt the epidemic (as the reproduction number remained above one), but that they were indeed effective in reducing the epidemic size [65, 66]. Estimates of the average duration of the infectious period of infected flocks were also similar between various models, e.g. 6.4 days [71], 7.47 days (95% CI: 7.2-7.8) [67] and 7.3 days (95% CI: 3.4-11.1) [65]. Although including spatial heterogeneity did not seem to have a substantial impact on the estimation of parameters such as the reproduction number and the average duration of the infectious period, spatial models had the additional ability to produce at-risk maps [67] and to evaluate spatially-explicit control strategies such as preventive culling around infected flocks [66, 70, 71].

Various ways of accounting for spatial heterogeneity in transmission during the 2003 HPAI H7N7 epidemics have been used. Le Menach et al. [66] estimated three transmission rates depending on the distance between farms and found substantially lower transmission rates for long-distance ranges (above 10 km) than for short- and medium-ranges (less than 3 km and between 3 and 10 km, respectively – Figure 3). Boender et al. [67] used a parametric transmission kernel with a logistic functional form to represent decreasing probability of transmission with increasing between-farm distance (Figure 3). The estimates of the spatial kernel parameters indicated a rapid decrease of the probability of transmission with increasing distance, where the probability was already halved at 1.9 km and reached very low values beyond 10 km (Figure 3; [67]). Finally, Seymour et al. [71] developed a non-parametric transmission kernel to represent the same phenomenon but in a more flexible way. The estimated non-parametric kernels were of similar shape and scale than the parametric kernel of Boender et al. [67], but with a higher degree of uncertainty, especially at lower distances [71]. The authors therefore argued that the assumptions imposed by parametric kernels may underestimate the uncertainty, and that non-parametric kernels may actually be more reflective of the actual uncertainty in the data [71].

The last point of comparison is the evaluation of control strategies, and more specifically of PEC. Boender et al. [67] concluded that culling farms in a 1-km radius was not very effective in reducing the basic reproduction number, but that culling farms in a 5-km radius reduced the basic reproduction number of all farms below one and thus was fully effective. On the contrary, Backer et al. [70] and Seymour et al. [71] both found a strong effect of PEC within a 1-km radius on the number of infected farms compared to no PEC. This difference may be explained by the way the control strategies were evaluated: in Boender et al. [67], the effectiveness of PEC was evaluated on the basic reproduction number of all farms, whereas in the other two studies, they simulated the epidemics under different scenarios and compared the model outputs, such as the number of infected farms and the number of culled farms. Therefore, the criteria used to evaluate the effectiveness of control strategies need to be carefully chosen. Interestingly, although Seymour et al. [71] found that increasing the culling radius beyond 1 km did not substantially reduce the number of infected farms but resulted in larger numbers of culled farms, Backer et al. [70] found that the effectiveness increased when expanding the radius from 1 to 3 km, but not from 3 to 10 km. This difference in the optimal PEC radius (1 vs 3 km) could be explained by the different culling capacity used, up to 20 farms per day in Backer et al. [70] *vs* only up to 6 farms per day in Seymour et al. [71], highlighting the importance of this parameter on the evaluation of control strategies. Overall, these results also emphasize the importance of comparing several models with different structures and assumptions to inform policy recommendations [82, 83].

**Model limitations and future avenues for modeling AIV transmission**

*Research gaps on avian influenza transmission and control*

It is worth noting that, until the winter of 2021 – 2022, the HPAI H5N8 subtype was responsible for the largest epidemics ever recorded in the EU in terms of the number of poultry outbreaks, geographical extent, and the number of dead wild birds, spreading across more than 29 European countries in the winters of 2016 – 2017 and of 2020 – 2021 [84]. However, only three studies in our review focused on parameter estimation and evaluation of control strategies of HPAI H5N8 in the EU [59–61]. Given the animal welfare and economic impact that these epidemics had, further research is needed on mechanistic modeling of HPAI H5N8 to better understand how the virus spread and how best to control it [28]. This subtype was replaced by the HPAI H5N1 since 2021, resulting in thousands of outbreaks in domestic poultry [84, 85]. Modeling studies will therefore also be required for this new devastating and widespread epidemic. In particular, evaluation of vaccination in an European context would be paramount given that it may become part of the EU strategy to mitigate the impact of HPAI epidemics [13].

Only six studies (13%) in our review included data on LPAI viruses, including four studies on the LPAI H7N9 epidemics that was first discovered in China in 2013. Given the zoonotic potential of these viruses and the ability of H5 and H7 LPAI viruses to mutate into HPAI, further research is needed on this topic [28]. This is especially true as each virus subtype and context is unique, and therefore a better understanding of the transmission of LPAI viruses in different settings is needed.

*Fine-tuning models*

Models are created for a specific situation with three key (and often opposing) characteristics to be balanced: accuracy, transparency, and adaptability [86]. The trade-off between them is inherent to the question being answered and the data availability. Even when the data is available, simplification is often necessary, for example to get a more adaptable model or save computational time. In doing so, some processes that could impact transmission dynamics may be overlooked.

This was the case in the reviewed articles for the impact of host species and production systems. Many models (25/45, 56%) considered an homogeneous "poultry" population. However, as was demonstrated in some of the included articles, galliformes and palmipeds contributed differently to different epidemics [32, 58, 60], and transmission parameters may vary between species depending on the virus subtype [28, 32, 58, 60] and between commercial and backyard farms [64, 68]. Thus, caution should be taken in pooling all the species as one population. Whenever possible and according to the available data, models should account for heterogeneity between host species and production systems, to get a better understanding of the contribution of each poultry populations to AIV transmission and design control strategies accordingly. Even if host heterogeneity is not included in the model, future studies should at least report information on the poultry species, e.g. to clarify which parameter values are applicable to which species [28].

A second common simplification was related to transmission routes. Most articles considered explicitely transmission between commercial poultry farms only, with only a few articles including explicit transmission from wildlife, backyard farms, or the environment [32, 41, 45, 48, 64, 68, 73, 74]. However, even if those sources of transmission may play a less important role relative to transmission between commercial farms, accounting for them may be important to accurately quantify the control efforts necessary to bring an epidemic under control [64].

Another modeling assumption that should be refined is the homogeneous mixing of epidemiological units, especially at the national [35, 37, 39, 44, 47, 49, 50] and global scales [40, 41]. This assumption bluntly means that every host interacts uniformly and randomly with every other hosts with the same probability, thus neglecting any heterogeneity that may arise from differences in age, behavior, and most importantly, spatial location [86]. This assumption helps simplify models in the absence of detailed data, and may be appropriate at small scales, such as for within-farm epidemic dynamics [31, 54, 55, 59, 61, 63, 69]. However, it may not adequately reflect transmission dynamics at larger scales. To refine this assumption and introduce spatial heterogeneity, spatial kernels were sometimes included at the between-farm and between-administrative units transmission scales (Figure 3). This may represent a simple but adequate alternative that is more representative of the spatial spread of AIV than homogeneous mixing.

Finally, the scale of the epidemiological unit was also important. Eight articles considered PBMs or IBMs using administrative areas as the epidemiological units [32, 33, 36, 45–47, 50, 52]. These administrative areas ranged from villages or communes, up to US counties or Nigerian states. In all these articles, the administrative area was considered as a single unit, i.e. the whole area became exposed or infectious at the same time, neglecting within-area dynamics. Rorres et al. [52] compared models using three different epidemiological units in Pennsylvania (US): farms, ZIP-codes and counties. Interestingly, the epidemic dynamics were consistent with those at the farm level when aggregating at the ZIP-code level, but not at the county level. These results suggest that counties are too large to neglect the within-area transmission dynamics [52]. Similar results were found when producing risk maps in The Netherlands: similar risk maps were obtained at the farm- or municipality-level, but not at the regional-level [27]. Therefore, modelling AIV transmission using epidemiological units larger than towns/cities should be avoided in the future.

*Estimating parameters from field data*

In most studies, only the transmisson parameter was estimated using field data (Table 3). Other parameters, such as the frequently used average durations of the latent and infectious periods, were most often assumed (Additional files 5 and 7).

For instance, most within-farm model parameters were assumed based on experimental studies for the same or different virus subtype. However, the course of the infection could be different in real-world situations compared to experimental studies, and could also differ between virus subtypes. Therefore, further insights could be gained on within-farm transmission dynamics and on the course of infection in individual birds by estimating parameters such as the durations of the latent and infectious periods from real-world epidemic data. Moreover, results from experimental studies could still be used as *a priori* knowledge, if the estimation is performed in a Bayesian framework [61]. The introduction time of the virus is another parameter that was estimated only in a few studies, despite its importance for contact tracing and identifying risk factors for the introduction of AIV in a farm [59]. Thus, using modeling approaches and real-world epidemic data to estimate this parameter represents a potential avenue for future research. Moreover, this parameter could be used to inform the time between the onset of infection to its detection (incubation period) in between-farm transmission models.

Indeed, the duration of the incubation period at the farm level was estimated in only three of the 23 articles considering between-farm transmission. Thus, results from within-farm modeling could be used to better inform this parameter and how it varies between farms according to their characteristics (species, size…). Alternatively, this parameter could be estimated as was done in Kim et al. [38], Hill et al. [42] and Andronico et al. [60]. This would provide useful information on the effectiveness of surveillance by farmers and veterinary services.

*Improving the evaluation of control strategies*

Given the frequency of devastating epidemics and the potential switch towards endemic circulation of HPAI viruses worldwide, preparedness and prevention strategies are paramount for the sustainability of poultry production. Mechanistic models of AIV transmission are of great value, since they provide the possibility to compare several strategies in a given epidemiological setting, which is rarely feasible in practice. Future evaluation of control strategies using mechanistic models could include comparison between different combinations of control strategies. Indeed, beside the combination of various strategies with the classically-used reactive culling, most articles compared strategies in isolation. However, the repeated occurrence of devastating HPAI H5Nx epidemics throughout the world raises serious concerns about the capacity of existing strategies to control AIV and stresses that no strategy would be sufficient by itself [14]. For instance, investigating the use of vaccination in combination with other control strategies could provide useful insights. The controversy in vaccine usage in Europe seems to be fading, especially with the catastrophic impact of the recent epidemic of HPAI H5NI during the winter of 2021 − 2022 [13]. The Netherlands and France, among others, already started clinical trials of vaccines [13]. Models can help address the effectiveness of control strategies including vaccination and the ability of veterinary services to implement such strategies promptly. The implementation of other mitigation approaches, such as reinforcing biosecurity measures and reducing the density of poultry farms, is a prerequisite to successful vaccination campaigns [14]. In this context, mechanistic models offer a promising route to decipher effects of associated mitigation and control strategies.

Furthermore, more is to gain from investigating the economic costs of implementing control strategies. Indeed, effective control strategies from an epidemiological point of view do not necessarily translate into efficient strategy from an economic point of view. In this review, only three articles evaluated the economic costs of an epidemic under various scenarios of control strategies [34, 70, 71]. Further work is therefore urgently needed on this topic.

While there are no strategies that can be used in every context, future investigations could be done with optimization algorithms that incorporate the parameters defining the mitigation strategies as input to determine the set of strategies to attain the desired outcomes (e.g., minimize the overall epidemic impact) [87]. This approach would replace the traditional method of comparing a set of predefined control strategies. For instance, instead of comparing a finite set of radiuses for PEC (e.g., 1, 3, 5 km), this method would find the optimal radius minimizing e.g. the epidemic impact or the economic cost. Such optimization approaches were for example used to determine the cost-effective surveillance strategies for invasive species management in Australia [88] or to choose the optimal allocation of biosecurity resources between quarantine and surveillance for protecting islands from pest invasion [87].

Real-time modeling could also bring useful insights to decision makers during an epidemic by comparing scenarios in the early stages of an epidemic [89]. However, limited data and resulting uncertainty in parameter values may limit the ability of models to provide useful insights. In this review, only one article retrospectively assessed the ability of its model to be used in real-time, by evaluating interventions using models fitted to data available at different time points during the epidemic [34]. Interestingly, despite high uncertainty in model projections, the ranking of control measures was consistent early on during the course of the epidemic (even when using only the first 10 days of the epidemics) when the objective was to reduce the epidemic size or the number of culled flocks. Similar results were found for two epidemics of foot-and-mouth diseases [89], supporting the ability of models to provide accurate comparisons of interventions relatively early during an epidemic. However, when the objective was to reduce the epidemic duration, using more data (25 days) was necessary to get consistent rankings, meaning that the recommendations to policy makers could be wrong beforehand [34]. Further research is therefore needed to assess the models' ability to be used in real-time for the evaluation of control strategies.

Finally, we identified in our review the importance of accounting for limited management capacity in mechanistic models of AIV. Indeed, not accounting for limited resources may overestimate the efficacy of interventions [43, 46, 58, 60, 66, 70, 71], which may lead to suboptimal or even wrong recommandations to policy makers. However, it may be difficult to quantify culling or vaccination capacity and how it changes over the course of an epidemic. Different assumptions regarding this parameter may lead to different conclusions, as seen for the 2003 H7N7 epidemics in The Netherlands [70, 71]. How to accurately model limited management capacity and how it changes over the course of an epidemic are therefore an avenue for future research. Collaboration with veterinary services and policy makers will be crucial to make realistic assumptions about this parameter, among others.

## DECLARATIONS

**Ethics approval and consent to participate:** Not applicable.

**Consent for publication:** Not applicable.

**Availability of data and materials:** All data generated or analysed during this study are included in this published article and its additional files.

**Competing interests:** The authors declare that they have no competing interests.

**Funding:** This work was funded by the FEDER/Région Occitanie Recherche et Sociétés 2018—AI-TRACK. This study was performed in the framework of the "Chair for Avian Health and Biosecurity", hosted by the National Veterinary College of Toulouse and funded by the Direction Générale de l'Alimentation, Ministère de l'Agriculture et de l'Alimentation, France. The funders had no role in the design of the study; in the collection, analyses, or interpretation of data; in the writing of the manuscript, or in the decision to publish the results.

**Authors' contributions:** TV and MP conceived the study design. BB performed the literature search. BB and AM performed the two-step screening, with ad hoc input from SL. SL, BB and AM extracted and analysed the data, and drafted the first version of the manuscript. All authors critically reviewed and edited each version of the paper, and approved the final manuscript.

**Acknowledgements:** Not applicable.

# TABLES

**Table 1. Overview of the 45 articles included in the review.**

| Subtype | Pathotype | Country | Epidemic years | References |
|---|---|---|---|---|
| H5N1 | HP | Thailand | 2004 – 2005 | Tiensin et al. [31]; Walker et al. [32]; Marquetoux et al. [33]; Retkute et al. [34] |
| | | | 2004 – 2007 | Van den Broek [35] |
| | | Vietnam | 2004 – 2005, 2007 | Walker et al. [36] |
| | | | 2008 – 2015 | Delabouglise et al. [37] |
| | | South Korea | 2008 | Kim et al. [38] |
| | | | 2003 – 2004, 2008, 2010 – 2011 | Kim and Cho [39] |
| | | Indonesia and global | 2008 – 2011 | Smirnova and Tuncer [40] |
| | | Global | 2005 – 2009 | Tuncer and Martcheva [41] |
| | | Bangladesh | 2007 – 2012 | Hill et al. [42]; Ssematimba et al. [44] |
| | | | 2007 – 2008, 2011 | Hill et al. [43] |
| | | India | 2008 – 2010 | Pandit et al. [45] |
| | | Nigeria | 2006 – 2007 | Pelletier et al. [46] |
| | | | 2005 – 2008 | Bett et al. [47] |
| | | | 2006 | Brown et al. [48] |
| | | Egypt | 2010 | Lee et al. [49] |
| | | Romania | 2005 – 2006 | Ward et al. [50] |
| H5N2 | HP | United States | 1983 – 1984 | Rorres et al. [51, 52] |
| | | | 2015 | Bonney et al. [53]; Ssematimba et al. [54] |
| | LP | | 2018 | Bonney et al. [55] |
| H5N6 | HP | Philippines | 2017 | Lee and Lao [56]; Salvador et al. [57] |
| | | South Korea | 2016 | Lee et al. [58] |
| | | | 2016 – 2018 | Kim and Cho [39] |
| H5N8 | HP | South Korea | 2016 | Lee et al. [58] |
| | | | 2014 – 2015, 2016 – 2018 | Kim and Cho [39] |
| | | Netherlands | 2014, 2016 | Hobbelen et al. [59] |
| | | France | 2016 – 2017 | Andronico et al. [60] |
| | | | 2020 – 2021 | Vergne et al. [61] |
| H7N1 | HP | Italy | 1999 – 2000 | Dorigatti et al. [62] |
| H7N3 | LP | Netherlands | 2003 | Gonzales et al. [63] |
| | HP | Canada | 2004 | Smith and Dunipace [64] |
| H7N7 | HP | Netherlands | 2003 | Stegeman et al. [65]; Le Menach et al. [66]; Boender et al. [67]; Bavinck et al. [68]; Bos et al. [69]; Backer et al. [70]; Seymour et al. [71] |
| H7N9 | LP | China | 2013 – 2015 | Li et al. [72] |
| | LP and HP | | 2013 – 2017 | Chen and Wen [73]; Bai et al. [74]; Zhu et al. [75] |

Box 1: definitions (see [86, 90–92])

**Population-based models:** although the overall population is made up of individual units, population-based models group all individual units of the same state together, without distinction between individual units belonging to the same subgroup. The number (or proportion) of the epidemiological units in each subgroup (e.g. susceptible, infectious, recovered) is tracked, but not the individual state of each unit (for instance, we know how many individual units are infectious, but not which ones). These models are also called compartmental models.

**Individual-based models:** contrary to population-based models, individual-based models monitor explicitly the state of each individual unit in the overall population. Therefore, it is possible to track both which individual units are in each state, and also the number of epidemiological units in each state. Individual-based models are useful for instance when individual characteristics cannot be describe in a discrete way (e.g. spatial location of the unit).

**Metapopulation models:** in metapopulation models, the overall population is subdivided into distinct (spatial) subpopulations. The model tracks transmission dynamics within each subpopulation, as well as between different subpopulations. Usually, interactions between units belonging to the same subpopulation are more frequent than interactions between units of different subpopulations. Metapopulation models can either track the number of individual units in each state, or monitor explicitly the state of each individual unit.

**Epidemiological unit:** the unit of interest and the smallest entity of the model. It could be an invididual animal, a group of animals, herds, or populations in regions or countries. The epidemiological unit can be aggregated and modeled as a number or proportion of the overall population in each state, or modeled as individuals whose status is tracked.

**Frequency-dependence:** with this assumption, the contact rate (the number of contacts made by each epidemiological unit per unit time, where contacts are of an appropriate type for transmission to be possible) is constant irrespective of the population size $N(t)$. In that case, the force of infection is of the form $\beta \frac{I(t)}{N(t)}$, where $\beta$ is the effective contact rate (in time$^{-1}$), i.e. the number of effective contacts made by each epidemiological unit per unit time. An effective contact is a contact that would effectively lead to transmission if the contact is between an infectious and a susceptible unit.

**Density-dependence:** with this assumption, the contact rate increases with the population size $N(t)$. In that case, the force of infection is of the form $\beta I(t)$, where $\beta$ (in unit$^{-1}$.time$^{-1}$) is the *per capita* rate at which two specific epidemiological units come into effective contact per unit time. Note that the density-dependent $\beta$ is not equivalent to the frequency-dependent $\beta$ because of these different assumptions on the contact rate.

**Table 2. The epidemiological unit and geographical scales found in the different model paradigms included in the review.**

| Model paradigm | Epidemiological unit | Geographical scales | | | | TOTAL |
|---|---|---|---|---|---|---|
| | | Global | National | Subnational | Local (farm) | |
| **Population-based model** | Administrative unit | - | 2 [47, 50] | 2 [33, 45] | - | 4 |
| | Farm | - | 4 [35, 37, 39, 44] | 3 [64, 65, 68] | - | 7 |
| | Individual birds | 2 [40, 41] | 1 [49, 56] | 5 [72–75] | 7 [31, 54, 55, 59, 61, 63, 69] | 15 |
| | TOTAL | 2 | 7 | 10 | 7 | 26 |
| **Individual-based model** | Administrative unit | - | 1 [46] | 3 [32, 36, 52] | - | 4 |
| | Farm | - | 4 [38, 46, 67, 71] | 9 [42, 43, 51–53, 57, 60, 62, 66] | - | 13 |
| | TOTAL | - | 4* | 11* | | 15* |
| **Metapopulation model** | Farm | - | 1 [34] | 1 [58] | - | 2 |
| | Individual birds | - | 2 [48, 70] | - | - | 2 |
| | TOTAL | - | 3 | 1 | - | 4 |

\* The total of individual-based models at the national and subnational scales is not the sum of individual-based models with administrative or farm units because Rorres et al. [52] and Pelletier et al. [46] used both types of epidemiological units.

**Table 3. Estimated parameter values in within- and between-farm transmission models.**

| Reference | Subtype | Model | Transmission rate $\beta$ | $R_0$ |
|---|---|---|---|---|
| **Within-farm transmission** | | | | |
| Tiensin et al. [31] | H5N1 HP | PBM/FD | Min: 0.60 (0.43-0.84) <br> Max: 2.30 (1.92-2.76) | Min: 2.18 (1.94-2.46) <br> Max: 3.49 (2.70-4.50) |
| Ssematimba et al. [54] | H5N2 HP | PBM/FD | 3.2 (2.3-4.3) | 12.8 (9.2-17.2) |
| Bonney et al. [55] | H5N2 LP | PBM/FD | Min: 0.5 (0.4-1.0) <br> Max: 6.0 (1.2-5.8) | - |
| Hobbelen et al. [59] | H5N8 HP | PBM/FD | Min: 0.95 (0.3-2.3) <br> Max: 34.4 (27.3-44.1) | - |
| Vergne et al. [61] | H5N8 HP | PBM/FD | 4.1 (2.8-5.8) | 17.5 (9.4-29.3) |
| Gonzales et al. [63] | H7N3 LP | PBM/FD | Min: 0.50 (0.45-0.55) <br> Max: 0.72 (0.68-0.77) | Min: 4.7 (3.0-8.6) <br> Max: 5.6 (4.3-7.7) |
| Bos et al. [69] | H7N7 HP | PBM/FD | 4.50 (2.68-7.57) | - |
| Backer et al. [70] | H7N7 HP | MPM/FD | 1.9 (0.61-8.1) | 7.6 |
| **Between-farm transmission** | | | | |
| Retkute et al. [34] | H5N1 HP | MPM/DD | 0.99 (0.76-1.12) $\times 10^{-6}$ | - |
| Delabouglise et al. [37] | H5N1 HP | PBM/DD | Min: 1.4 $\times 10^{-8}$ – Max: 40 $\times 10^{-8}$ | Min: 0.55 – Max: 15.7 |
| Kim et al. [38] | H5N1 HP | IBM/DD | 0.429 (probability) | - |
| Kim and Cho [39] | H5N1 HP <br> H5N8 HP <br> H5N6 HP | PBM/DD | - | Min: 0.03 (0-0.98) <br> Max: 2.20 (1.51-3.16) |
| Hill et al. [42, 43] | H5N1 HP | IBM/DD | Min: 1.71 (0.586-3.63) $\times 10^{-10}$ <br> Max: 1.06 (0.0729-3.78) $\times 10^{-7}$ | - |
| Ssematimba et al. [44] | H5N1 HP | PBM/FD | Min: 0.08 (0.06-0.10) <br> Max: 0.11 (0.08-0.20) | Min: 0.85 (0.77-1.02) <br> Max: 0.96 (0.72-1.20) |
| Bonney et al. [53] | H5N2 HP | IBM/DD | 0.0061 (0.0025-0.0137) | - |
| Salvador et al. [57] | H5N6 HP | IBM/DD | 0.0012 (0.0001-0.1) | - |
| Lee et al. [58] | H5N6 HP <br> H5N8 HP | MPM/DD | Min: 0.00007 – Max: 0.00707 | 1.3427 |
| Andronico et al. [60] | H5N8 HP | IBM/FD | Min: 0.23 (0.16-0.31) <br> Max: 0.53 (0.37-0.72) | - |
| Dorigatti et al. [62] | H7N1 HP | IBM/DD | Min: 0.0009 (0.0005-0.0013) <br> Max: 0.0155 (0.0078-0.0232) | - |
| Smith and Dunipace [64] | H7N3 HP | PBM/DD | Min: 0 – Max: 0.00238 | 4.8 |
| Stegeman et al. [65] | H7N7 HP | PBM/DD* | Min: 0.17 (0.1-0.2) <br> Max: 0.47 (0.3-0.7) | Min: 1.2 (0.6-1.9) <br> Max: 6.5 (3.1-9.9) |
| Le Menach et al. [66] | H7N7 HP | IBM/DD | Min: 0.076 – Max: 0.336 | Min: 1.5 (1-2.5) <br> Max: 5.2 (4-6.9) |
| Boender et al. [67] | H7N7 HP | IBM/DD | 0.002 (0.0012-0.0039) | - |
| Bavinck et al. [68] | H7N7 HP | PBM/DD | 1.7 (1.5-2.0) $\times 10^{-4}$ | 1.33 |
| Backer et al. [70] | H7N7 HP | MPM/DD | 0.0039 (0.0023-0.0076) | - |

FD: frequency-dependence ($\beta$ is in day$^{-1}$); DD: density-dependence ($\beta$ is in farm$^{-1}$.day$^{-1}$).

\* This model assumed a force of infection of the form $\beta \frac{I(t)}{N}$, where $N$ was a constant (the initial population size). Although assuming a density-dependent contact rate [65], the unit of $\beta$ was the same as for frequency-dependent models, here in day$^{-1}$.

**Table 4. Mitigation strategies evaluated in articles included in the review.**

| Reference | Outcome for comparison[*] | Mitigation strategies evaluated[#] | | | | |
|---|---|---|---|---|---|---|
| | | RC | PEC | Vacc | Surv | Other |
| Andronico et al. [60] | Epi | x | x | | x | |
| Backer et al. [70] | Epi + Eco | | x | x | | |
| Boender et al. [67] | Epi | | x | | | |
| Bonney et al. [53] | Epi | | | | | Early marketing |
| Brown et al. [48] | Epi | x | | | x | Quarantine and movement control |
| Chen and Wen [73] | Epi H | | | x | | Biosecurity measures |
| Dorigatti et al. [62] | Epi | | x | | | Ban on restocking |
| Hill et al. [43] | Epi | | x | x | x | |
| Kim et al. [38] | Epi | | x | | x | |
| Le Menach et al. [66] | Epi | x | x | | x | National movement ban |
| Lee and Lao [56] | Epi | | | x | | Quarantine |
| Lee et al. [49] | Epi H | x | | | | |
| Lee et al. [58] | Epi | | x | | | |
| Pandit et al. [45] | Epi | x | | | | |
| Pelletier et al. [46] | Epi | | x | x | | |
| Retkute et al. [34] | Epi + Eco | | x | x | | |
| Salvador et al. [57] | Epi | | x | | x | |
| Seymour et al. [71] | Epi + Eco | | x | | | |
| Walker at al. [36] | Epi | | | x | x | |
| Walker et al. [32] | Epi | | | | x | |
| Zhu et al. [75] | Epi H | | | | | Biosecurity measures and closure of live poultry markets |

[*] Outcome considered by the authors to compare mitigation strategies: Epi, epidemiological parameters in poultry; Eco, economical impact of measures evaluated; Epi H: epidemiological parameters in humans.
[#] Mitigation strategies evaluated: RC, reactive culling; PEC, pre-emptive culling; Vacc, vaccination; Surv, surveillance.

**FIGURES**

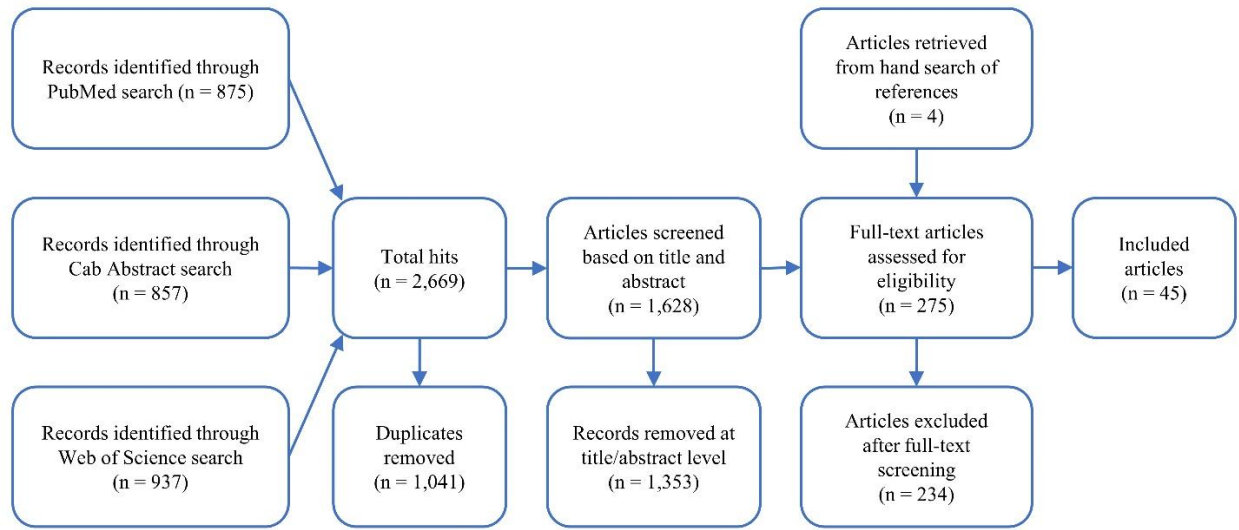

**Figure 1. PRISMA flow diagram of the selection process for including articles in the review**.

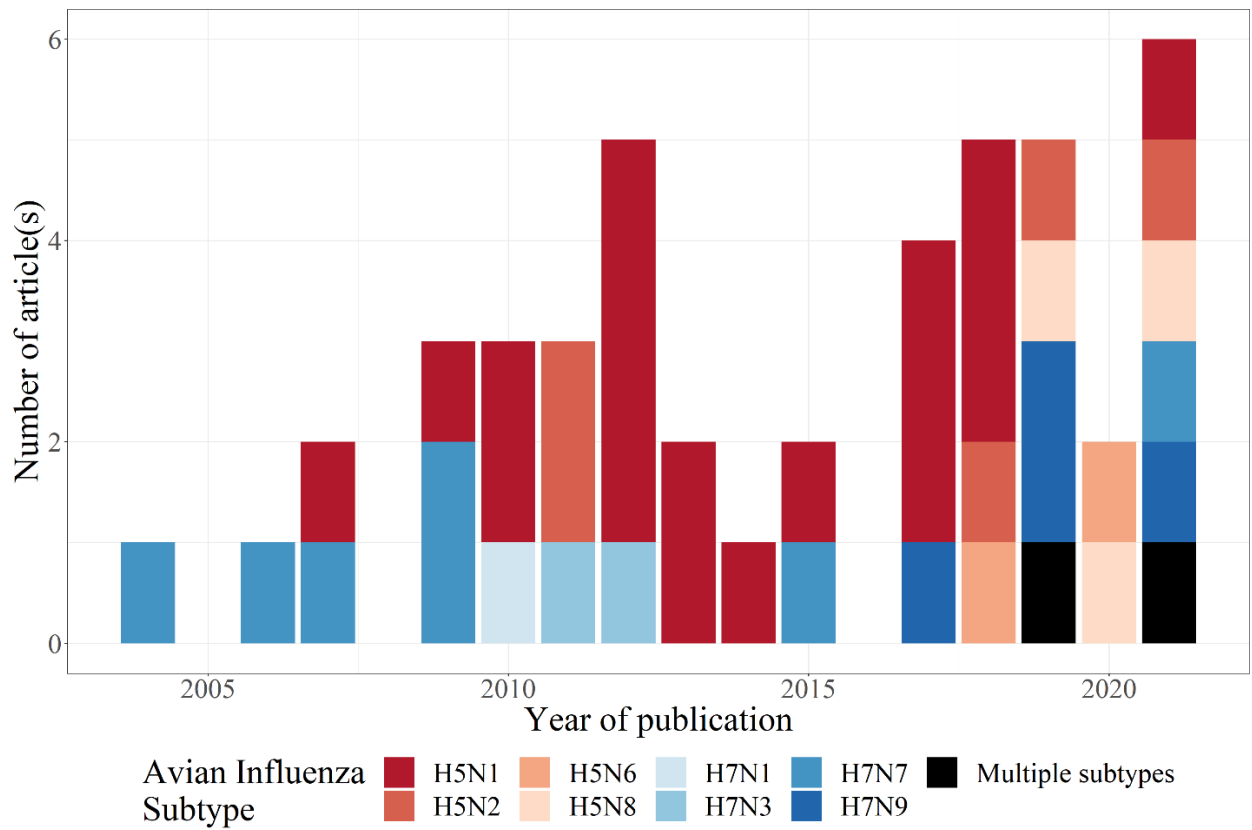

**Figure 2. Temporal description - Distribution of publication date for different avian influenza subtypes.**
Both Lee et al. [58] and Kim and Cho [39] studied multiple subtypes circulating in South Korea.

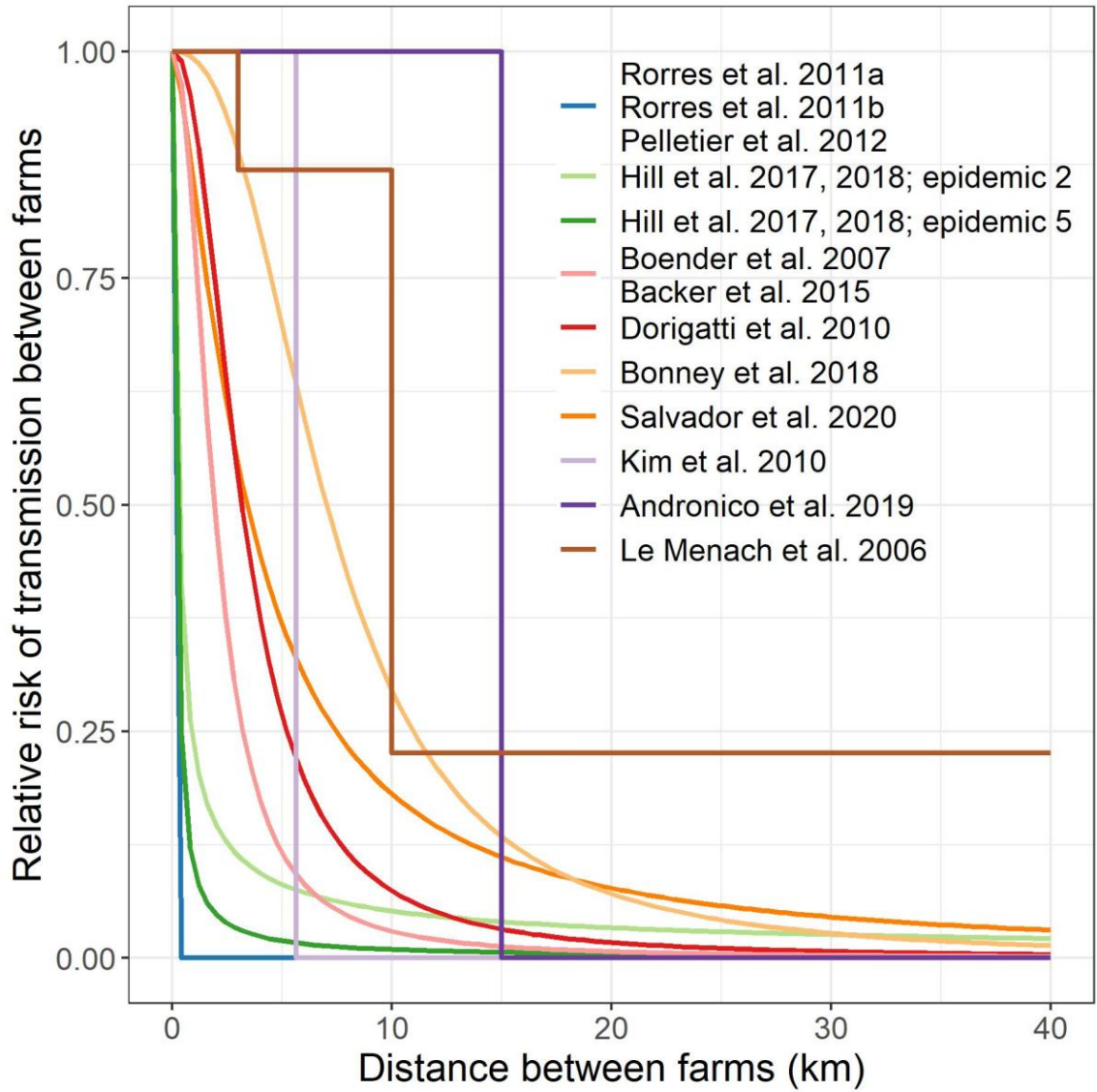

**Figure 3. Comparison of spatial transmission kernels used in 13 articles modeling between-farm transmission of HPAI.**
Note that for Rorres et al. [51, 52] and Pelletier et al. [46], althought the relative risk of transmission between farms decreased to very low values at small distances, the absolute probability of transmission remained substantial even at long distances because the size of both susceptible and infectious farms were included in the force of infection.

**ADDITIONAL FILES**

**Additional file 1: Complete overview of search terms used in PubMed, Web of Science and CAB Abstracts.**

**Additional file 2: Summary of the two-step screening.**

**Additional file 3: Supporting data.** [31–75]

**Additional file 4: References of the 45 included articles.** [31–75]

**Additional file 5: Parameter values used in within-farm transmission models.** [31, 54, 55, 59, 61, 63, 69, 70, 93–127]

**Additional file 6: Parameter values used in between-farms transmission models.** [34, 35, 37–39, 42–44, 46, 51–53, 57, 58, 60, 62, 64–68, 70, 71, 95, 96, 115, 121, 124, 127–134]

**Additional file 7: Comparison of spatial transmission kernel estimates used in articles studying between-farm transmission.** [42, 43, 46, 51–53, 57, 62, 67, 70]